\documentclass[11pt]{article}
\usepackage{amssymb}
\author{Keith Anguige\footnote{E-mail address: anguige@aei-potsdam.mpg.de}\\~\\Max Planck Institut fuer Gravitationsphysik\\Albert Einstein Institut, Am Muehlenberg 1\\14476 Golm, Germany}
\title{A class of plane symmetric perfect-fluid cosmologies with a Kasner-like singularity}
\begin{document}
\maketitle
\begin{abstract}
We prove the existence of a class of plane-symmetric perfect fluid cosmologies with a Kasner-like singularity of cigar type. These solutions of the Einstein equations depend on two smooth functions of one space coordinate. They are constructed by solving a symmetric hyperbolic system of Fuchsian equations.

\end{abstract}

\section{Introduction}
The Bianchi I spacetimes are the simplest class of anisotropic cosmological models. They admit a 3-parameter abelian group of isometries acting on spacelike hypersurfaces \cite{xv}. The dynamics of solutions to Einstein's equations with this symmetry type are particularly well-understood in the case of a perfect fluid as matter source. For a physically reasonable, non-stiff equation of state and with an appropriate choice of time-orientation all solutions isotropise in the future and in the past any solution which is not FRW must become Kasner-like as the initial singularity is approached. Solutions with all values of the Kasner exponents are realised \cite{x}.
 
Under the additional assumptions  of local rotational symmetry and reflection symmetry much can also be said about the asymptotic behaviour of collisionless gas cosmologies (i.e. solutions of the Einstein-Vlasov equations). Once again all solutions isotropise in the future. For any solution which is not FRW the asymptotic generalised Kasner exponents (i.e. the eigenvalues of the second fundamental form divided by the mean curvature) in the past are either~$(-\frac{1}{3},\frac{2}{3},\frac{2}{3})$~or~$(0,\frac{1}{2},\frac{1}{2})$~\cite{xiii}. The first case is Kasner like  and is generic in that it occurs for an open dense set of initial data. Note that no solution of the ~$(0,\frac{1}{2},\frac{1}{2})$~type occurs in the fluid case and that the Kasner exponents~$(1,0,0)$~are not realised with the collisionless gas.

It is natural to ask whether the above qualitative features are maintained if one relaxes the requirement of spatial homogeneity somewhat. Certainly there are many inhomogeneous FRW-like solutions of both the Einstein-perfect fluid equations and the Einstein-Vlasov equations, as has been shown in the framework of isotropic singularites \cite{i}-\cite{iii}, \cite{vi}, \cite{vii}. Kasner like solutions of the Einstein-Vlasov equations have been investigated by Rein \cite{ix} under the assumption of plane symmetry, which is the simplest inhomogeneous generalisation of LRS Bianchi I symmetry. By proving a global existence theorem he is able to show that an open set of~$C^{\infty}$~data on a regular Cauchy surface evolves back to a Kasner like singularity of the~$(-\frac{1}{3},\frac{2}{3},\frac{2}{3})$~type, agreeing with the picture described in \cite{x}. In the present paper we study plane-symmetric, non-stiff~$\gamma$-law perfect fluid cosmologies with asymptotic geometry of this same type. The results obtained can be thought of as providing some hard evidence for certain aspects of the general picture of cosmological singularities proposed by Belinskii, Khalatnikov and Lifschitz (BKL) many years ago \cite{xviii}. This picture of the generic solution of Einstein's equations, based on formal calculations, has a number of key elements:
\begin{itemize}
\item[(i)] 
Near a singularity the evolution at different spatial points should decouple so that spacetime is approximated by a spatially homogeneous model at each point.
\item[(ii)]
The matter content of spacetime should in general not have a significant effect on the dynamics near the singularity. Thus the dynamics should be well approximated by vacuum spacetimes.

\item[(iii)]
The most general spatially homogeneous cosmology is modelled by a spacetime of Bianchi type VIII or IX and vacuum spacetimes of Bianchi types VIII and IX typically exhibit Mixmaster behaviour.
\end{itemize}

One should note that there are exceptional models which do not fit the BKL picture. Spacetimes with isotropic singularities are an example, but they depend on fewer free functions than the general solution and so do not constitute a counterexample. Also, while the BKL picture predicts in general complicated, oscillatory Mixmaster behaviour one may still investigate (i) and (ii) for less complicated Bianchi type I-like models. Rigorous results for Gowdy spacetimes have been obtained by Isenberg and Moncrief \cite{xvii} and Kichenassamy and Rendall \cite{v}. Recall that the Gowdy spacetimes are defined as spatially compact solutions of the vacuum Einstein equations which have two commuting Killing vectors and in which the so-called `twist constants' vanish. In \cite{v} the so-called Fuchsian method is used to construct a class of  Gowdy spacetimes which have Kasner-like singularities and depend on the maximum number of free-functions. In \cite{xvii} the authors specialize to so-called polarized Gowdy symmetry and use energy estimates to show that all solutions are `Asymptotically Velocity Term Dominated' near a cosmological singularity. This is one way of saying that the field equations spatially decouple as in (i).

The method used in this paper to investigate perfect-fluid spacetimes is very similar to that employed in \cite{v}. It consists in taking a two-parameter family of exact Bianchi I solutions, making an inhomogeneous perturbation and solving a singular (Fuchsian) system of field equations for the perturbation near the singularity~`$t=0$'. In this way we construct a two free-function family of~$C^{\infty}$~inhomogeneous cosmologies with the required asymptotic Bianchi I like behaviour. This family is `generic' in the sense that the general plane-symmetric perfect-fluid spacetime depends on just two free-functions.

The main result is stated formally as Theorem 6.1 at the end of section 6.

In section 7 we show that the spacetimes constructed have crushing singularities.

\section{Exact solutions}
The general Bianchi I  solution of the Einstein-perfect fluid equations, as given in \cite{xiv} is
\begin{equation}ds^{2}= -B^{2(\gamma-1)}d\tau^{2}+\tau^{2p_{1}}B^{2q_{1}}dx^{2}+\tau^{2p_{2}}B^{2q_{2}}dy^{2}+\tau^{2p_{3}}B^{2q_{3}}dz^{2}\end{equation}

where

\begin{equation} B^{2-\gamma}=\alpha+m^{2}\tau^{2-\gamma}~~~~~~\alpha \geq 0,~m>0\end{equation}
\begin{equation}p_{1}+p_{2}+p_{3}=1~~,~~~p_{1}^{2}+p_{2}^{2}+p_{3}^{2}=1~~,~~~q_{i}=\frac{2}{3}-p_{i}\end{equation}
(The case~$\alpha=0$~is an FRW model and is excluded in what follows.)

The density of the fluid is given by
\begin{equation}\mu=\frac{4m^{2}}{3\tau^{\gamma}B^{\gamma}}\end{equation}
and~$\gamma$~is the polytropic index, so that the pressure is given by~$P=(\gamma-1)\mu$.

It is a consequence of the momentum constraint that in Bianchi type I the fluid cannot tilt and hence the 4-velocity points in the direction of~$\frac{\partial}{\partial\tau}$.

Now if one wants to consider the plane-symmetric cases of (1) then there are two choices for the Kasner exponents~$p_{i}$, namely~$(-\frac{1}{3}, \frac{2}{3}, \frac{2}{3})$~and~$(1, 0, 0)$. In this paper we will restrict attention to the~$(-\frac{1}{3}, \frac{2}{3}, \frac{2}{3})$~case and consider inhomogeneous perturbations of the metric which depend on a coordinate in the~`$-\frac{1}{3}$~direction'.

When we come to write down Einstein's equations for a plane symmetric metric we will use conformal coordinates, i.e. coordinates for which the metric takes the form

\begin{equation}ds^{2}= e^{2A(t, x)}(-dt^{2}+ dx^{2}) + R(t, x)(dy^{2}+dz^{2})\end{equation}

The model solution (1) in the case~$p_{i}=(-\frac{1}{3}, \frac{2}{3}, \frac{2}{3})$~may be written in conformal coordinates by making the transformation
\begin{equation}\tau\rightarrow t=\int_{0}^{\tau}\frac{s^{1/3}}{\alpha+m^{2}s^{2-\gamma}}~ds\end{equation}
Then the metric (1) takes the form (5) with
\begin{equation}A=-\frac{1}{3}\log{\tau}+\frac{1}{2-\gamma}\log{(\alpha+m^{2}\tau^{2-\gamma})}~~,~~R=\tau^{4/3}\end{equation}
and~$\tau$~given implicitly as a function of~$t$~by the relation (6).

\section{The Einstein-perfect fluid equations}
We will assume that spacetime is filled with a polytropic perfect fluid, so that the stress tensor takes the form
\begin{equation}T^{\alpha\beta}=(\rho +P)u^{\alpha}u^{\beta}+Pg^{\alpha\beta}\end{equation}
with~$u^{\alpha}u_{\alpha}=-1$~and~$P=(\gamma-1)\rho,~1<\gamma<2$.

With this stress tensor the Einstein evolution equations for the metric (5) are
\begin{equation}R_{tt}-R_{xx}= (2-\gamma)Re^{2A}\rho\end{equation}
\begin{equation}A_{tt}-A_{xx}-\frac{1}{4}R^{-2}(R_{t}^{2}-R_{x}^{2})=-\frac{1}{2}\gamma e^{2A}\rho\end{equation}
while the constraints read
\begin{equation}R^{-1}R_{xx}-\frac{1}{4}R^{-2}(R_{t}^{2}+R_{x}^{2})-R^{-1}R_{t}A_{t}-R^{-1}R_{x}A_{x}=-e^{2A}\rho(\gamma (v^{1})^{2}+1)\end{equation}
\begin{equation}-R^{-1}R_{tx}+R^{-1}(R_{t}A_{x}+A_{t}R_{x})+\frac{1}{2}R^{-2}R_{t}R_{x}=-e^{2A}\gamma\rho v^{0}v^{1}\end{equation}
where~$v^{i}=e^{A}u^{i}$~and hence~$(v^{0})^{2}-(v^{1})^{2}=1$.

The Euler equations~$\nabla^{\alpha}T_{\alpha\beta}=0$~take the following explicit form
\begin{displaymath}(1+\gamma(v^{1})^{2})\rho_{t}+2\gamma\rho (v^{1})_{t}-2\gamma\rho A_{t}(1+(v^{1})^{2})+\gamma v^{0}v^{1}\rho_{x}\end{displaymath}
\begin{displaymath}+\gamma\rho\left\{\left(\frac{(v^{1})^{2}+(v^{0})^{2}}{v^{0}}\right)((v^{1})_{x}-A_{x}v^{1})-\frac{v^{1}}{v^{0}}A_{x}\right\}\end{displaymath}
\begin{equation}+\gamma\rho\left\{A_{t}(3(v^{0})^{2}+(v^{1})^{2})+v^{0}v^{1}(4A_{x}+R^{-1}R_{x})+(v^{0})^{2}R^{-1}R_{t}\right\}=0\end{equation}
and
\begin{displaymath}\gamma v^{1}v^{0}\rho_{t}+\gamma\rho\left\{\left(\frac{(v^{1})^{2}+(v^{0})^{2}}{v^{0}}\right)((v^{1})_{t}-A_{t}v^{1})-\frac{v^{1}}{v^{0}}A_{t}\right\}\end{displaymath}
\begin{displaymath}+\left(\gamma(1+(v^{1})^{2})-1\right)\rho_{x}+2\gamma\rho v^{1}(v^{1})_{x}\end{displaymath}
\begin{equation}+\gamma\rho\left\{v^{1}v^{0}(2A_{t}+R^{-1}R_{t})+(v^{1})^{2}R^{-1}R_{x}+(1+2(v^{1})^{2})A_{x}+2v^{0}v^{1}A_{t}\right\}=0\end{equation}

\section{Inhomogeneous perturbations}

We now choose two strictly positive smooth functions~$\alpha(x),~m(x)$~and seek solutions~$\rho,~v^{1},~A,~R$~of (9)-(14) which approach the model forms (4)-(7)~as~$t\rightarrow 0$.\\
To be specific we make the following ansatz
\begin{equation}A=-\frac{1}{3}\log{\tau}+\frac{1}{2-\gamma}\log{(\alpha(x)+m^{2}(x)\tau^{2-\gamma})}+t\tilde{A}\end{equation}
\begin{equation}R=\tau^{4/3}(1+\tilde{R})\end{equation}
\begin{equation}\log{\rho}=\log{\mu}+t^{\frac{3}{4}(\gamma-\frac{2}{3})+\epsilon}\phi\end{equation}
\begin{equation}v^{1}=t^{\frac{3}{4}(\gamma-\frac{2}{3})}(\psi^{0}+t^{\epsilon}\tilde{\psi})\end{equation}
where~$0<\epsilon <\frac{3}{4}(2-\gamma)$,~$\tau (t, x)$~is given by (6) and
\begin{displaymath}\psi^{0}\equiv \left(\frac{4}{3}\alpha\right)^{\frac{3}{4}(\gamma-\frac{2}{3})}(m^{2}\gamma)^{-1}\left(\frac{\gamma-1}{\gamma-2}\right)\frac{\alpha_{x}}{\alpha}\end{displaymath}

The aim is to find solutions of the Einstein-perfect fluid equations for which\linebreak[4]$\tilde{A},~\tilde{R},~\phi$~and~$\tilde{\psi}$~tend to zero as~$t$~tends to zero. The choice of~$\psi^{0}$~ensures that the momentum constraint is satisfied at~$t=0$~(see section 6).

It is convenient to write the field equations for~$\tilde{A},~\tilde{R},~\phi$~and~$\tilde{\psi}$~in first order form. To do this we introduce the following new variables:
\begin{displaymath}U=\tilde{R}_{t},~~Q=\partial_{t}(t\tilde{A}),~~X=\tilde{R}_{x},~~Y=t\tilde{A}_{x},~~S=t^{-1}\tilde{R}\end{displaymath}

In terms of these variables the evolution equations (9)-(10) take the form
\begin{equation}tX_{t}=tU_{x}\end{equation}
\begin{equation}tY_{t}=tQ_{x}\end{equation}
\begin{equation}t\tilde{R}_{t}=tU\end{equation}
\begin{equation}t\tilde{A}_{t}+\tilde{A}-Q=0\end{equation}
\begin{equation}t\partial_{t}S+S-U=0\end{equation}
\begin{displaymath}tU_{t}+2U=tX_{x}+\left(-2t^{\frac{3}{4}(2-\gamma)}F(t, x)-\frac{8}{3}m^{2}t\tau^{\frac{2}{3}-\gamma}\right)U\end{displaymath}
\begin{displaymath}+\frac{t}{\tau^{4/3}}(\tau^{4/3})_{xx}(1+\tilde{R})+2t(\log{\tau^{4/3}})_{x}X\end{displaymath}
\begin{equation}+\frac{4m^{2}}{3}t\tau^{-\frac{2}{3}-\gamma}(1+\tilde{R})(2-\gamma)(\alpha+m^{2}\tau^{2-\gamma})(e^{2t\tilde{A}+t^{1+\epsilon -p}\phi}-1)\end{equation}
\begin{displaymath}tQ_{t}-\frac{1}{2}U=tY_{x}+tA^{0}_{xx}-\frac{2}{3}\gamma m^{2}t\tau^{-\frac{2}{3}-\gamma}(\alpha+m^{2}\tau^{2-\gamma})(e^{2t\tilde{A}+t^{1+\epsilon -p}\phi}-1)\end{displaymath}
\begin{displaymath}+\frac{1}{4}t(1+\tilde{R})^{-2}\left(U^{2}+\frac{8}{3}\tau^{\frac{2}{3}-\gamma}m^{2}(1+\tilde{R})U+((\log{\tau^{4/3}})_{x}(1+\tilde{R})+X)^{2}\right)\end{displaymath}
\begin{equation}+\frac{1}{2}U\left\{t^{\frac{3}{4}(2-\gamma)}F(t, x)+t(1+t^{\frac{3}{4}(2-\gamma)}F(t, x))S\right\}\end{equation}
where~$A^{0}$~stands for the first two terms in (15), the function~$F(t, x)$, continuous in~$t$~and smooth in~$x$~is given by the relation
\begin{equation}\frac{4}{3}\alpha t\tau^{-\frac{4}{3}}=1+t^{\frac{3}{4}(2-\gamma)}F(t, x)\end{equation}

The Euler equations may, after some rearrangement, be written
\begin{displaymath}\left(1+\gamma t^{2-2p}\psi^{2}+\frac{2\gamma t^{2-2p}\psi^{2}(v^{0})^{2}}{1+2t^{2-2p}\psi^{2}}\right)t^{\epsilon -p}(t\phi_{t}+(1+\epsilon -p)\phi)=-\gamma t^{1-p}(v^{0})^{-1}(1+2t^{2-2p}\psi^{2})\psi_{x}\end{displaymath}
\begin{displaymath}+\gamma t^{1-p}(v^{0})^{-1}(1+2t^{2-2p}\psi^{2})\psi(A^{0}_{x}+Y)(v^{0})^{-1}(1+v^{0})+\gamma^{2}t^{2-2p}\psi^{2}\tau^{-4/3}(\alpha+m^{2}\tau^{2-\gamma})\end{displaymath}
\begin{displaymath}+\frac{2\gamma t^{1-p}\psi v^{0}}{1+2t^{2-2p}\psi^{2}}\left\{\left(t^{2-2p}\psi^{2}+\frac{\gamma-1}{\gamma}\right)((\log{\mu})_{x}+t^{1+\epsilon -p}\phi_{x})-\gamma t^{1-p}\psi v^{0}\tau^{-4/3}(\alpha+m^{2}t^{2-\gamma})\right.\end{displaymath}
\begin{displaymath}+2t^{2-2p}\psi\psi_{x}+(1+2t^{2-2p}\psi^{2})(A^{0}_{x}+Y)+(\log{\tau^{4/3}})_{x}\end{displaymath}
\begin{displaymath}\left.+(1+\tilde{R})^{-1}X+t^{1-p}\psi v^{0}(4\tau^{\frac{2}{3}-\gamma}+4Q+(1+\tilde{R})^{-1}U\right\}\end{displaymath}
\begin{displaymath}-\gamma\left\{Q+\frac{2}{3}t^{2-2p}\psi^{2}(\alpha\tau^{-4/3}+4m^{2}\tau^{2-\gamma}+3Q)+4v^{0}t^{1-p}\psi(A^{0}_{x}+Y)\right.\end{displaymath}
\begin{equation}\left.+\left(\frac{1+t^{2-2p}\psi^{2}}{1+\tilde{R}}\right)U+v^{0}t^{1-p}\psi\left((\log{\tau^{4/3}})_{x}+\frac{X}{1+\tilde{R}}\right)+t^{1-p}\psi v^{0}((\log{\mu})_{x}+t^{1+\epsilon -p}\phi_{x})\right\}\end{equation}
and
\begin{displaymath}\left(\frac{1+2t^{2-2p}\psi^{2}}{v^{0}}\right)t^{\epsilon -p}(t\tilde{\psi}_{t}+\epsilon\tilde{\psi})=-\left(t^{2-2p}\psi^{2}+\frac{\gamma-1}{\gamma}\right)((\log{\mu})_{x}+t^{1+\epsilon -p}\phi_{x})\end{displaymath}
\begin{displaymath}+t^{1-p}\psi\left(1+\frac{1}{v^{0}}\right)\left(\frac{2}{3}m^{2}\tau^{\frac{2}{3}-\gamma}+Q\right)-\frac{\psi}{4}F\left(1+\frac{1}{v^{0}}\right)+\frac{\psi}{4}t^{2-3p}\psi^{2}\frac{(1+t^{2}\psi^{2-2p})^{-\frac{1}{2}}}{(1+\sqrt{1+t^{2-2p}\psi^{2}})}\end{displaymath}
\begin{displaymath}-t^{1-p}\psi v^{0}\left\{4m^{2}\tau^{\frac{2}{3}-\gamma}+\frac{U}{1+\tilde{R}}+4Q+(\log{\tau^{4/3}})_{x}+\frac{X}{1+\tilde{R}}\right\}\end{displaymath}
\begin{displaymath}+\frac{t^{1-p}\psi v^{0}}{(1+\gamma t^{2-2p}\psi^{2})}\left\{\gamma t^{1-p}\left(\frac{1+2t^{2-2p}\psi^{2}}{v^{0}}\right)(\psi_{x}-\psi(A^{0}_{x}+Y)(1+v^{0})(v^{0})^{-1})\right.\end{displaymath}
\begin{displaymath}-\gamma^{2}t^{2-2p}\psi^{2}\tau^{-\frac{4}{3}}(\alpha+m^{2}\tau^{2-\gamma})+\gamma t^{1-p}\psi(2t^{1-p}\psi_{t}+(2-2p)t^{-p}\psi+v^{0}((\log{\mu})_{x}+t^{1+\epsilon -p}\phi_{x}))\end{displaymath}
\begin{displaymath}+\gamma\left(Q+2t^{2-2p}\psi^{2}(-\frac{1}{3}\alpha\tau^{-\frac{4}{3}}+\frac{2}{3}m^{2}\tau^{\frac{2}{3}-\gamma}+Q)+4v^{0}t^{1-p}\psi(A^{0}_{x}+Y)\right)\end{displaymath}
\begin{displaymath}\left.+\gamma\left(\left(\frac{1+t^{2-2p}\psi^{2}}{1+\tilde{R}}\right)U+\frac{4}{3}\tau^{-\frac{4}{3}}t^{2-2p}\psi^{2}(\alpha+m^{2}\tau^{2-\gamma})+v^{0}t^{1-p}\psi\left((\log{\tau^{\frac{4}{3}}})_{x}+\frac{X}{1+\tilde{R}}\right)\right)\right\}\end{displaymath}
\begin{displaymath}+\frac{3\gamma}{4}\left(2(1+t^{p}F)v^{0}\alpha^{-1}m^{2}\tau^{2-\gamma}t^{-p}\psi+F(1+t^{2-2p}\psi^{2})^{\frac{1}{2}}\psi+\frac{t^{2-3p}\psi^{3}}{(1+t^{2-2p}\psi^{2})^{\frac{1}{2}}+1}\right)\end{displaymath}
\begin{equation}-2\gamma\psi t^{2-2p}\psi_{x}-(1+2t^{2-2p}\psi^{2})(A^{0}_{x}+Y)\end{equation}
where~$p=\frac{3}{4}(2-\gamma)$~and~$\psi=\psi^{0}+t^{\epsilon}\tilde{\psi}$.

\section{Existence and uniqueness of solutions}
The Einstein evolution equations (19)-(25) take the form
\begin{equation}ta_{1}(w_{1})_{t}+tb_{1}(w_{1})_{x}+c_{1}w_{1}=t^{\delta}G_{1}\left(w_{1}, \phi, t, x\right)\end{equation}
where~$w_{1}$~stands for~$U, \tilde{A}, Q, X, Y, \tilde{R}, S$~,~$a_{1}$~is the~$6\times 6$~identity matrix,~$b_{1}$~is a constant symmetric matrix and~$c_{1}$~is a constant 
matrix with eigenvalues~$0, 1, 2$.~$G_{1}$~is a vector-valued function, continuous in~$t$~and smooth in all other arguments.~$\delta$~is a strictly positive constant.

The Euler equations (27)-(28) may be symmetrised by multiplying one of the equations through by the smooth function of~$t^{1-p}\psi$~given by
\begin{equation}\left(\frac{4\gamma t^{2-2p}\psi^{2}v^{0}}{1+2t^{2-2p}\psi^{2}}-\frac{\gamma(1+2t^{2-2p}\psi^{2})}{v^{0}}\right)\left(\frac{1-\gamma}{\gamma}+\frac{\gamma\psi^{2}t^{2-2p}(v^{0})^{2}}{1+\gamma t^{2-2p}\psi^{2}}-t^{2-2p}\psi^{2}\right)^{-1}\end{equation}

In this way (27)-(28) give rise to a system of the form
\begin{equation}ta_{2}(t^{1-p}\psi)(w_{2})_{t}+tb_{2}(\psi, t)(w_{2})_{x}+c_{2}w_{2}=t^{\delta}G_{2}(w_{1}, w_{2}, t, x)\end{equation}
where~$w_{2}$~stands for~$\phi,~\tilde{\psi}$,~$a_{2}$~is a diagonal positive definite matrix, smooth as a function of~$t^{1-p}\psi$~(and equal to the identity at~$t^{1-p}\psi=0$),~$b_{2}$~is a smooth symmetric matrix and~$c_{2}$~is diagonal with positive eigenvalues.~$G_{2}$~is a vector-valued function, continuous in~$t$~and smooth in its other arguments.

Combining (29)-(31) we see that the field equations may be expressed as
\begin{equation}ta_{3}(t^{1-p}\psi)(w_{3})_{t}+tb_{3}(\psi, t)(w_{3})_{x}+c_{3}w_{3}=t^{\delta}G_{3}\left(w_{3}, t, x,\right)\end{equation}
where~$a_{3}$~is smooth and positive definite,~$b_{3}$~is smooth and symmetric and~$c_{3}$~is constant 
with positive eigenvalues.~$G_{3}$~is a vector-valued function, continuous in~$t$~and smooth in all other arguments.

Now note that we may choose~$\delta=\frac{1}{n}$~and~$\frac{3}{4}(\gamma-2/3)=q/n$~for some positive integer~$n$~and some~$q>1$~wlog. With this choice of~$\delta$~it is then convenient to make the change of time coordinate~$t\rightarrow s=t^{\frac{1}{n}}$. Equation (32) then takes the form
\begin{equation}\tilde{a}_{3}(w_{3})_{s}+\tilde{b}_{3}(w_{3})_{x}+\frac{1}{s}c_{3}w_{3}=G_{3}\end{equation}
where~$\tilde{a}_{3}=n^{-1}a_{3}$~and~$\tilde{b}_{3}=s^{n-1}b_{3}$.

It is also convenient to multiply the right hand side of (33) by a smooth cut-off function of~$x$~with compact support. Call the result~$\tilde{G}_{3}$.

We are aiming to solve the equation
\begin{equation}\tilde{a}_{3}(w_{3})_{s}+\tilde{b}_{3}(w_{3})_{x}+\frac{1}{s}c_{3}w_{3}=\tilde{G}_{3}\end{equation}
with~$w_{3}(0)=0$~on~$[0, T]\times\mathbb{R}$. A solution of this equation will be a local (in space) solution of (33) by the finite speed of propagation.

For the application of energy estimates to a Fuchsian system it is important that the singular part of the system be positive definite. Unfortunately the matrix~$c_{3}$~in (34) is not positive definite even though its eigenvalues are all positive. The situation can however be remedied by making a change of variables. First, because~$c_{3}$~has positive eigenvalues one can, by Lemma 2.1 of \cite{xx} obtain an approximate solution~$u_{i}$~of (34) which satifies this equation up to ~$O(s^{i})$~for any positive integer~$i$, i.e.
\begin{displaymath}s\tilde{a}_{3}(u_{i})_{s}+s\tilde{b}_{3}(u_{i})_{x}+c_{3}u_{i}-s\tilde{G}_{3}= O(s^{i})\end{displaymath}
as~$s\rightarrow 0$. The function~$u_{i}(s, x)$~is continuous in~$s$~at~$s=0$~and smooth for~$s>0$, with~$u_{i}(0, x)=0$.

Now define the new variable~$\tilde{w}_{3}$~according to~$\tilde{w}_{3}=s^{-i}(w_{3}-u_{i})$. Then as an equation for~$\tilde{w}_{3}$~(34) becomes
\begin{displaymath}\tilde{a}_{3}(\tilde{w}_{3})_{s}+\tilde{b}_{3}(\tilde{w}_{3})_{x}+\frac{1}{s}\tilde{c}_{3}\tilde{w}_{3}=\hat{G}_{3}\end{displaymath}
where~$\tilde{c}_{3}=c_{3}+iI$~and~$\hat{G}_{3}$~has the same properties as~$\tilde{G}_{3}$.

We now assume that~$i$~was chosen large enough to make~$\tilde{c}_{3}$~positive definite and in the following we drop the tildes from~$\tilde{w}_{3}$~and~$\tilde{c}_{3}$~for convenience.

Now~$\hat{G}_{3}$~contains terms like~$f(t, x)g(w_{3})w_{3}$~or just~$h(t, x)$~where~$f$~and~$h$~are continuous in~$t$~and smooth with compact support in~$x$~and~$g$~is smooth. We approximate such functions~$f, h$~uniformly in the Sobolev space~$H^{p}(\mathbb{R})$~on~$[0, T]$~by sequences of smooth ($C^{\infty}$) functions of compact support~$f^{n}, h^{n}$~(see Lemma A.1 of the Appendix and see \cite{xvi} for an introduction to Sobolev spaces).

Consider the equation
\begin{equation}\tilde{a}_{3}(s,w_{3}^{n})(w_{3}^{n+1})_{s}+\tilde{b}_{3}(s,w_{3}^{n})(w_{3}^{n+1})_{x}+\frac{1}{s}c_{3}w_{3}^{n+1}=\sum_{i} f_{i}^{n}g_{i}(w_{3}^{n})w_{3}^{n+1}+h_{i}^{n}\end{equation}
where the right hand side represents an approximation to~$\hat{G}_{3}$. Since this equation is linear, all coefficients are smooth and~$c_{3}$~is positive definite, we can iteratively obtain smooth solutions~$w_{3}^{n}$~on~$[0, T]$~with~$w_{3}^{n}(0)=(0)$~by Theorem 9.1 of \cite{iv}.
\textbf{Lemma 5.1}~The sequence~$w_{3}^{n}$~is uniformly bounded in~$H^{p}$~and~for the element~$\tilde{\psi}^{n}$~of~$w_{3}^{n}$~we have that~$s\partial_{s}\tilde{\psi}^{n}$~is uniformly bounded in~$H^{p-1}$~on~$[0, T_{1}]$~for some~$T_{1}\leq T$. (~$p$~is chosen large enough to apply the Sobolev embedding lemma \cite{xvi} wherever needed).

\textit{Proof}~We use induction.\\
Suppose\\
(i)~$\|w_{3}^{n}\|_{p}(s)\leq R$ for~$s\leq T$\\
(ii)~$\|s\partial_{s}\tilde{\psi}^{n}\|_{p-1}\leq L$~for~$s\leq T$

Apply~$x$~derivatives of order~$\alpha$~to (35) to get
\begin{equation}\tilde{a}_{3}(w_{3}^{n})\partial_{s}\nabla^{\alpha}w_{3}^{n+1}+\tilde{b}_{3}^{n+1}\partial_{x}\nabla^{\alpha}w_{3}^{n+1}+\frac{1}{s}c_{3}\nabla^{\alpha}w_{3}^{n+1}=M^{\alpha}\end{equation}
where~$M^{\alpha}$~contains terms like
\begin{equation}\tilde{a}_{3}\nabla^{\alpha}\left(\tilde{a}_{3}^{-1}\tilde{b}_{3}\partial_{x}w_{3}^{n+1}\right)\end{equation}
and the terms coming from the~$f_{i}^{n}$~and~$h_{i}^{n}$.

Now take the inner product of (36) with~$\nabla^{\alpha}w_{3}^{n+1}$~and use the symmetry of~$\tilde{b}_{3}$, the boundedness of~$s\partial_{s}\tilde{\psi}^{n}$~and the uniform equivalence of the~$L^{2}$~norm of~$\nabla^{\alpha}w_{3}^{n+1}$~to
\begin{equation}\int\tilde{a}_{3}\nabla^{\alpha}w_{3}^{n+1}\cdot\nabla^{\alpha}w_{3}^{n+1}~dx\end{equation}
in a standard way \cite{viii}, to get the following inequality
\begin{equation}\|w_{3}^{n+1}\|_{p}(s)\leq K_{1}\int_{0}^{s}\|w_{3}^{n+1}\|_{p}+\|w_{3}^{n}\|_{p}+K_{2}~dr\end{equation}
(We used the fact that the~$f^{n}$~are bounded in~$H^{p}$ and we also used the positive definiteness of~$c_{3}$~to throw away the singular~$\frac{1}{s}$~terms which occur in (36))\\
If~$T_{1}$~is now chosen sufficiently small Gronwall's inequality gives~$\|w_{3}^{n+1}\|_{p}(s)\leq R$~for~$s\leq T_{1}$.

From the differential equation (35) one may easily use a Moser estimate \cite{viii} to obtain~$\|s\partial_{s}\tilde{\psi}^{n+1}\|_{p-1}\leq L$~for~$s\leq T_{1}$~if~$L$~was chosen large enough. Hence the lemma.

Note: From the form of~$\tilde{a}_{3}$~the bound on~$s\partial_{s}\tilde{\psi}^{n}$~is enough to give control over the time derivative of~$\tilde{a}_{3}(w_{3}^{n})$~which occurs in the right hand side of the energy estimate for~$w_{3}^{n+1}$.

\textbf{Lemma 5.2}~The sequence~$w_{3}^{n}$~converges uniformly in~$L^{2}$~on$~[0, T_{2}]$~for some~$T_{2}\leq T_{1}$.

\textit{Proof.} By subtracting the  equation satisfied by~$w_{3}^{n}$~from that satisfied by~$w_{3}^{n-1}$~and again exploiting the positive definiteness of~$c_{3}$~one obtains the following~$L^{2}$~energy estimate
\begin{displaymath}\|w_{3}^{n}-w_{3}^{n-1}\|_{2}(s)\leq K\int_{0}^{s}\left\{\|w_{3}^{n}-w_{3}^{n-1}\|_{2}+\|w^{n-1}-w^{n-2}\|_{2}\right.\end{displaymath}
\begin{equation}+\left.\|f^{n-1}-f^{n-1}\|_{2}\right\}~dr\end{equation}
(Note; we need boundedness of the  quantity~$(\tilde{a}_{3}(w_{3}^{n})-\tilde{a}_{3}(w_{3}^{n-1}))\partial_{s}w_{3}^{n}$~to obtain this inequality. But this follows from the form of~$\tilde{a}_{3}$~and Lemma 5.1)
Put~$\alpha^{n}=\|w_{3}^{n}-w_{3}^{n-1}\|_{2},~~\gamma^{n}=\|f^{n-1}-f^{n-2}\|_{2}$. Then we have
\begin{equation}\alpha^{n}(s)\leq K\int_{0}^{s}\left[\alpha^{n}+\alpha^{n-1}+\gamma^{n}\right]~dr\end{equation}
Gronwall implies
\begin{equation}\alpha^{n}\leq K\int_{0}^{s}\left[\alpha^{n-1}+\gamma^{n}\right]~dr\end{equation}
which implies
\begin{equation}\sup_{s\leq T_{2}}\alpha^{n}\leq K\int_{0}^{t}\sup_{s\leq T_{2}}\alpha^{n-1}+\gamma^{n}~dr\end{equation}
Thus
\begin{equation}\sup_{s\leq T_{2}}\alpha^{n}\leq KT_{2}\sup_{s\leq T_{2}}\alpha^{n-1}+K\sup_{s\leq T_{2}}\gamma^{n}\end{equation}
We assume that the sequence~$f_{n}$~was chosen to converge so fast that~$\sum\gamma^{n}<\infty$. We also choose~$T_{2}$~so that$~KT_{2}<1$. It now follows by the argument of Corollary 5.4 in \cite{viii} that~$w_{3}^{n}$~is uniformly Cauchy in~$L^{2}$~on~$[0, T_{2}]$. Hence the lemma.

The inequality of Gagliardo-Nirenberg \cite{viii} now shows that~$w_{3}^{n}$~converges uniformly in~$H^{p-1}$~to some~$w$~on~$[0, T_{2}]$.~$p$~was arbitrarily large. From the differential equation one see that quantities~$\partial_{s}w_{3}^{n}$~converge uniformly on~$(\epsilon , T_{2}]$~for~$\epsilon> 0$. Hence~$w$~is~$C^{1}$~by the Sobolev imbedding theorem. A standard continuation argument \cite{xii} for regular symmetric hyperbolic systems can be applied away from~$s=0$~to get that~$w$~is~$C^{\infty}$~for~$s>0$.

Suppose now that we have two solutions~$w,~\hat{w}$~of 
the field equations, smooth for~$s>0$, continuous at~$s=0$~and such that~$w(0)=\hat{w}(0)$. Then by taking the difference of the equations satisfied by~$w$~and~$\hat{w}$~one may obtain the following ~$L^{2}$~energy estimate.
\begin{equation}\partial_{s}\|w-\hat{w}\|_{2}^{2}\leq K\|w-\hat{w}\|_{2}^{2}\end{equation}
for~$s>0$.
(Again one needs control over$~(\tilde{a}_{3}(w)-\tilde{a}_{3}(\hat{w}))\partial_{s}w$, which follows from the equation)

Hence
\begin{equation}\|w-\hat{w}\|_{2}^{2}(s)-\|w-\hat{w}\|_{2}^{2}(\epsilon)\leq K\int_{\epsilon}^{s}\|w-\hat{w}\|_{2}^{2}~dt\end{equation}
which implies
\begin{equation}\|w-\hat{w}\|_{2}^{2}(s)\leq K\int_{0}^{s}\|w-\hat{w}\|_{2}^{2}~dt + \|w-\hat{w}\|_{2}^{2}(\epsilon)\end{equation}
Gronwall's inequality now implies
\begin{equation}\|w-\hat{w}\|_{2}^{2}(s)\leq e^{Ks}\|w-\hat{w}\|_{2}^{2}(\epsilon)\end{equation}
Since~$\epsilon >0$~was arbitrary, we must have~$w\equiv\hat{w}$.

\section{The constraints}
Define constraint quantities~$C_{0},~C_{1}$~by
\begin{equation}C_{0}=R^{-1}R_{xx}-\frac{1}{4}R^{-2}(R_{t}^{2}+R_{x}^{2})-R^{-1}R_{t}A_{t}-R^{-1}R_{x}A_{x}+e^{2A}\rho(1+\gamma (v^{1})^{2})\end{equation}
\begin{equation}C_{1}=-R^{-1}R_{tx}+R^{-1}(R_{t}A_{x}+A_{t}R_{x})+\frac{1}{2}R^{-2}R_{t}R_{x}+e^{2A}\gamma\rho v^{0}v^{1}\end{equation}
If the evolution equations (9)-(10), (13)-(14) are satisfied then a calculation shows that the following hold
\begin{equation}\partial_{t}C_{0}=-\partial_{x}C_{1}-\frac{R_{t}}{R}C_{0}-\frac{R_{x}}{R}C_{1}\end{equation}
\begin{equation}\partial_{t}C_{1}=-\partial_{x}C_{0}-\frac{R_{t}}{R}C_{1}-\frac{R_{x}}{R}C_{0}\end{equation}
One also calculates that the quantities~$\tilde{C_{0}}=tC_{0},~\tilde{C_{1}}=tC_{1}$~tend to zero as~$t$~tends to zero. These quantities satisfy the following
\begin{equation}t\partial_{t}\tilde{C_{0}}+\left(\frac{tR_{t}}{R}-1\right)\tilde{C_{0}}=-t\partial_{x}\tilde{C_{1}}-\frac{tR_{x}}{R}\tilde{C_{1}}\end{equation}
\begin{equation}t\partial_{t}\tilde{C_{1}}+\left(\frac{tR_{t}}{R}-1\right)\tilde{C_{1}}=-t\partial_{x}\tilde{C_{0}}-\frac{tR_{x}}{R}\tilde{C_{0}}\end{equation}
Now~$R_{x}/R$~is~$O(1)$~and~$tR_{t}/R=1+O(t^{\delta})$~for some~$\delta >0$. It thus follows from the ideas of section (5) that~$\tilde{C_{0}}$~and~$\tilde{C_{1}}$~are identically zero and hence the constraints are satisfied.

The results of sections 5 and 6 can now be summarised as follows:
\\\\
~~~~\textbf{Theorem 6.1}. Given two smooth, strictly positive functions~$\alpha(x),~m(x)$~on~$\mathbb{R}$\\there exists a unique solution~$(g_{\mu\nu}, \rho, u^{\alpha})$~of the Einstein equations coupled to a~$\gamma$-law perfect fluid on~$\mathbb{R}^{3}\times (0, T)$~satisfying
\begin{displaymath}ds^{2}=e^{2A(t, x)}(-dt^{2}+dx^{2})+R(t, x)(dy^{2}+dz^{2})\end{displaymath}
and
\begin{displaymath}A=A^{0}+o(t)\end{displaymath}
\begin{displaymath}R=\tau^{\frac{4}{3}}(1+o(1))\end{displaymath}
\begin{displaymath}\log{\rho}=\log{\mu}+o(t^{1-\frac{3}{4}(2-\gamma)+\epsilon})~~,~0<\epsilon <\frac{3}{4}(2-\gamma)\end{displaymath}
\begin{displaymath}v^{1}=o(t^{\frac{3}{4}(\gamma-2/3)})\end{displaymath}
\begin{displaymath}t\partial_{t}A=t\partial_{t}A^{0}+o(1)=-\frac{1}{4}+o(1)\end{displaymath}
\begin{displaymath}\partial_{t}R=\frac{4}{3}\alpha +o(t)\end{displaymath}
where~$1<\gamma<2$~and
\begin{equation}A^{0}=-\frac{1}{3}\log{\tau}+\frac{1}{2-\gamma}\log{(\alpha(x)+m^{2}(x)\tau^{2-\gamma})}\end{equation}
\begin{displaymath}t=\int_{0}^{\tau}\frac{s^{\frac{1}{3}}}{\alpha+m^{2}s^{2-\gamma}}ds\sim\frac{3}{4\alpha}\tau^{\frac{4}{3}}\end{displaymath}

\section{Crushing singularities}
The extrinsic curvature of the hypersurfaces of constant~$t$~is given by
\begin{equation}K=e^{-t\tilde{A}}\tau^{-1}(\alpha+m^{2}\tau^{2-\gamma})^{\frac{1}{\gamma-2}}(\alpha+2m^{2}\tau^{2-\gamma}+\tau^{4/3}(Q+U(1+\tilde{R})^{-1}))\end{equation}
It follows that in the limit as~$t\rightarrow 0$~we have~$K\sim\alpha^{(\gamma+2)/(\gamma-2)}t^{-\frac{3}{4}}$. Thus if~$\alpha$~is chosen uniformly positive we have a crushing singularity at~$t=0$. 
{
\section{Conclusion}
We have proved the existence of a family of plane-symmetric perfect-fluid cosmologies which have a~$(-\frac{1}{3},\frac{2}{3},\frac{2}{3})$~Kasner-like singularity and which depend on the full number of free functions for perfect-fluid spacetimes with this symmetry type, namely two. This  supports the idea of Belinskii, Khalatnikov and Lifschitz that the Einstein equations should generically decouple spatially near a cosmological singularity and that the matter content of spacetime should have a negligible effect on the dynamics near the singularity.

It would be interesting to know if the result obtained in this paper can be extended to perfect-fluid spacetimes with polarized Gowdy symmetry (i. e. diagonal~$G_{2}$~cosmologies). The formal expansions of BKL and the work of Isenberg and Moncrief suggest that this should indeed be the case. It may well be that the Fuchsian algorithm can once again be applied in this more general setting, but this has not yet been attempted.

\appendix
\section{Approximation of continuous functions by smooth functions}.

\textbf{Lemma A.1} Suppose we are given a function~$F(t, x)$~on~$[0,T]\times\mathbb{R}$, which is continuous in~$t$~and smooth in~$x$. Suppose also that~$F$~has fixed compact support in~$x$. Then there is a sequence of smooth functions~$F_{n}(t, x)~$on~$[0,T]\times\mathbb{R}$~such that~$F_{n}\rightarrow F~$in~$H^{s}(\mathbb{R})$, uniformly on~$[0,T]$.

\textit{Proof} First define
\begin{equation}\tilde{F}(t, x)=\left\{\begin{array}{cc}F(t, x)&0\leq t\leq T\\
F(0, x)&-1\leq t\leq 0\\F(T, x)&t\geq T\end{array}\right.\end{equation}
and let
\begin{equation}F_{n}(t, x)=\int_{\mathbb{R}}J_{\frac{1}{n}}(t-s)\tilde{F}(s, x)~ds\end{equation}
where~$J_{\epsilon}$~is a smooth function, supported on~$(-\epsilon, \epsilon)$, such that~$\int J_{\epsilon}(x)dx=1$.\\
Then
\begin{displaymath}|\tilde{F}(t, x)-F_{n}(t, x)|=\left|\int_{\mathbb{R}}J_{\frac{1}{n}}(t-s)(\tilde{F}(t, x)-\tilde{F}(s, x))~ds\right|\end{displaymath}
\begin{displaymath}\leq\sup_{|t-s|\leq\frac{1}{n}}|\tilde{F}(t, x)-\tilde{F}(s, x)|\end{displaymath}
Now
\begin{equation}\sup_{0\leq t\leq T}\int_{\mathbb{R}}|(\tilde{F}-F_{n})(t, x)|^{2}~dx\leq\int_{\mathbb{R}}\sup_{0\leq t\leq T}\left(\sup_{|t-s|\leq\frac{1}{n}}|\tilde{F}(t, x)-\tilde{F}(s, x)|\right)^{2}~dx\end{equation}
Clearly the integrand in the right hand side of (59) tends to zero for each~$x$~by the uniform continuity of~$\tilde{F}$~in~$t$. It is also clear that the sequence of integrands is bounded by an integrable function. Thus~$F_{n}\rightarrow F$~in~$L^{2}$~uniformly on~$[0,T]$~by Lebesgue's dominated convergence theorem. One may now apply the above argument to partial derivatives to get the desired result.

\end{document}